\documentclass[10pt, doublecolumn]{IEEEtran}
\usepackage{graphicx}
\usepackage{caption}
\usepackage[ruled]{algorithm2e}
\ifCLASSOPTIONcompsoc
  \usepackage[caption=false,font=normalsize,labelfont=sf,textfont=sf]{subfig}
\else
  \usepackage[caption=false,font=footnotesize]{subfig}
\fi
\usepackage{indentfirst}
\usepackage{graphicx}
\usepackage{bm}
\usepackage{amsmath}
\allowdisplaybreaks[4]
\usepackage{amssymb}
\usepackage{times}
\usepackage{mathtools}
\usepackage{psfrag}
\usepackage{cite}
\usepackage{lastpage}
\usepackage{fancyhdr}
\usepackage{color}
 \usepackage{amsthm}
\usepackage{bigints}
\newtheorem{Lemma}{Lemma}
\theoremstyle{remark}
\newtheorem{Remark}{$\quad$Remark}
\newtheorem{Corollary}{Corollary}

\newtheorem{corollary}[Corollary]{$\mathbf{Corollary}$}

\newcommand{\Equ}[1]{
  \begin{align}
    #1
  \end{align}}

\newcommand{\SubEquL}[2]{
  \begin{subequations}\label{#1}
    \begin{align}
     #2
    \end{align}
\end{subequations}}
\begin{document}
\title{On Energy Efficiency of Hybrid NOMA}
\author{Yanshi Sun, \IEEEmembership{Member, IEEE}, Zhiguo Ding, \IEEEmembership{Fellow, IEEE}, Yun Hou, \IEEEmembership{Senior Member, IEEE}, and George K. Karagiannidis, \IEEEmembership{Fellow, IEEE}
\thanks{Y. Sun is  with School of Computer Science and Information
Engineering, Hefei University of Technology, Hefei, 230009, China. (email: sys@hfut.edu.cn)
Z. Ding is with Department of Electrical Engineering and Computer
Science, Khalifa University, Abu Dhabi, UAE. (email: zhiguo.ding@ieee.org).
Y. Hou is with Department of Computer Science, Hang Seng University of Hong Kong, Hong Kong. (email: aileenhou@hsu.edu.hk).
G. K. Karagiannidis is with Department of Electrical and Computer
Engineering, Aristotle University of Thessaloniki, Greece and also with
Artificial Intelligence \& Cyber Systems Research Center, Lebanese American
University (LAU), Lebanon (geokarag@auth.gr).
}\vspace{-3em}}
\maketitle
\begin{abstract}
This paper aims to prove the significant superiority of hybrid non-orthogonal multiple access (NOMA) over orthogonal multiple access (OMA) in terms of energy efficiency. In particular, a novel hybrid NOMA scheme is proposed in which a user can transmit signals not only by using its own time slot but also by using the time slots of other users. The data rate maximization problem is studied by optimizing the power allocation, where closed-form solutions are obtained. Furthermore, the conditions under which hybrid NOMA can achieve a higher instantaneous data rate with less power consumption than OMA are obtained. It is proved that the probability that hybrid NOMA can achieve a higher instantaneous data rate with less power consumption than OMA approaches one in the high SNR regime, indicating the superiority of hybrid NOMA in terms of power efficiency. Numerical results are also provided to verify the developed analysis and also to demonstrate the superior performance of hybrid NOMA.
\end{abstract}
\begin{IEEEkeywords}
Hybrid NOMA, energy efficiency, power allocation, data rate maximization, performance analysis
\end{IEEEkeywords}

\section{Introduction}
The development of next-generation multiple access (NGMA) techniques to enable sixth-generation (6G) mobile communications has recently attracted considerable attention and effort from both academia and industry \cite{you2021towards}.
In particular, non-orthogonal multiple access (NOMA) has been recognized as an important candidate for NGMA \cite{liu2022evolution, makki2020survey, sun2023hybrid}. For example, NOMA is expected to be considered to meet the future requirements of the recently released International Mobile Telecommunications (IMT)-2030 Framework \cite{recommendation2023framework}.

Hybrid NOMA, which can be treated as a general form of NOMA and conventional orthogonal multiple access (OMA), has been proposed recently \cite{ding2022hybrid, ding2024design,ding2024hybrid}. Unlike existing schemes in \cite{shao2021resource,choi2019evolutionary}, where a user can choose to transmit in either OMA or NOMA, in hybrid NOMA, a user can divide its transmission into several orthogonal subchannels (which can be time slots \cite{ding2022hybrid}, subcarriers \cite{ding2024hybrid}, spatial beams \cite{ding2024design}, etc.), and in each subchannel, the user can transmit either in NOMA mode or OMA mode. The advantages of NOMA are mainly twofold \cite{ding2024hybrid}. One is its flexibility in implementation, since hybrid NOMA can be developed as a simple add-on to the existing OMA-based legacy network framework. The other is its effective resource utilization, since hybrid NOMA allows multidimensional resource allocation, which provides more freedom for optimization.

Although promising, research on hybrid NOMA is still in its infancy. This paper aims to investigate the superior advantage of hybrid NOMA in terms of energy efficiency by seeking the answer to the following fundamental question: can hybrid NOMA achieve a higher data rate while consuming less energy than OMA? The main contributions of this thesis are as follows:
\begin{itemize}
  \item A novel hybrid NOMA scheme is proposed, where a user can transmit signals not only by using its own timeslot, but also by using other users' timeslots. The instantaneous data rate maximization problem for the proposed hybrid NOMA scheme is solved with a closed-form solution.
  \item Based on the obtained expression for the achievable data rate of hybrid NOMA, the comparisons with OMA are made. Specifically, the conditions under which hybrid NOMA can achieve a higher instantaneous data rate with less energy consumption compared to pure OMA are provided. Furthermore, the probability of these conditions is analyzed taking into account the randomness of the channel gains. It is shown that the probability for hybrid NOMA to achieve a higher instantaneous data rate with less power consumption compared to OMA approaches one in the high SNR regime.
\end{itemize}

\section{System Model}
\begin{figure}[!t]
  \centering
      \setlength{\abovecaptionskip}{0em}   % ����ͼƬ������ͼƬ����
      \setlength{\belowcaptionskip}{-2em}   % ����ͼƬ���������ľ���
  \includegraphics[width=2.1in]{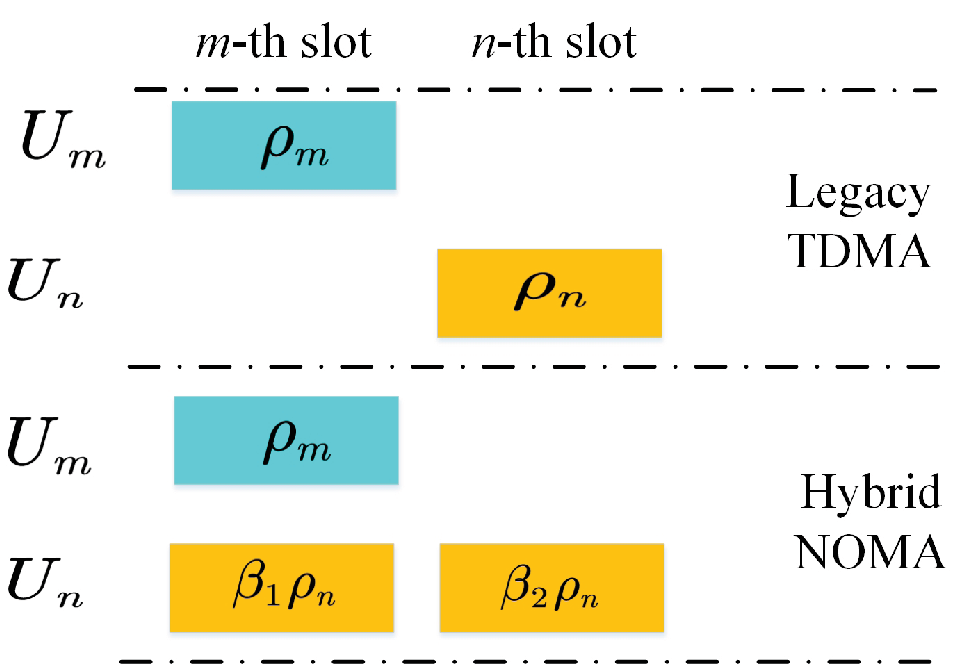}\\
  \caption{Illustration of the system model.}\label{system_model}
\end{figure}

Consider a legacy TDMA based uplink communication scenario with one base station, and two users which are denoted by
$U_m$ and $U_n$, respectively, as shown in Fig. \ref{system_model}. Each user is allocated with an individual time slot with duration $T$.
Thus the achievable instantaneous data rate of $U_v$ ($v\in\{m,n\}$) in TDMA is given by
\Equ{R_{v}^{\text{OMA}}=\log\left(1+|h_v|^2\rho_v\right),} where $h_v$ is the small-scale fading modeled as
a circularly symmetric complex gaussian (CSCG) random variable with mean zero and variance $1$, i.e.,
$h_v\sim \mathcal{CN}(0,1)$, and $\rho_v$ is the transmit power. It is worth noting that the additive background noise power is normalized in this paper.

It is assumed that $U_m$ is a quality of service (QoS) sensitive user which has a target data rate denoted by $R_0$. Specifically, when the achievable data rate $R_{m}^{\text{OMA}}$ is larger than $R_0$,
$U_m$ only wants to transmit with $R_0$ instead of a higher data rate; and when
$R_{m}^{\text{OMA}}\leq R_0$, $U_m$ would like to transmit with $R_{m}^{\text{OMA}}$ which is most
close to $R_0$ for reliable transmission. On the other hand, $U_n$ wants to transmit as much data as possible.
Based on the above assumptions, the considered hybrid NOMA scheme allows $U_n$ to have an additional
transmitting opportunity by transparently sharing $U_m$'s slot (NOMA transmission), in addition to its own dedicated
slot (OMA transmission). Note that the ``transparency'' means that the admission of $U_n$ to $U_m$'s slot should not
degrade the performance of $U_m$ compared to OMA.
The achievable rates of $U_n$ in the considered hybrid NOMA scheme for NOMA and OMA transmission are given by:
\begin{itemize}
\item for NOMA transmission, $U_n$ transmits signals with power $\beta_1 \rho_n$, where $\beta_1$ is the power coefficient, $0\leq\beta_1\leq1$. The BS first decodes $U_m$'s signal, and if succeeded, successive interference cancellation (SIC) will be carried out and $U_n$'s signal can be decoded without any interference, yielding the following data rate:
    \begin{align}
     R_{n}^1=\log\left(1+\beta_1 \rho_n |h_n|^2\right).
    \end{align}
To ensure the success of SIC, the transmit power of $U_n$ needs to satisfy $\beta_1 \rho_n |h_n|^2 \leq \tau_m$, where $\tau_m=\max\{0, \rho_m|h_m|^2/\epsilon_0-1\}$, $\epsilon_0=2^{R_0}-1$. Note that $\tau_m$ can be treated as the maximal interfering power with which $U_m$'s signal can be successfully decoded \cite{sun2021new}.
\item for OMA transmission, $U_n$ transmits signals with power $\beta_2 \rho_n$, where $\beta_2$ is the power coefficient, $0\leq\beta_2\leq1$, yielding the following data rate:
 \begin{align}
     R_{n}^2=\log\left(1+\beta_2 \rho_n |h_n|^2\right).
    \end{align}
\end{itemize}
Thus, the achievable rates of $U_n$ in the considered hybrid NOMA scheme can be expressed as
$R_n=R_{n}^1+R_{n}^2$.

\section{Rate maximization for hybrid NOMA and comparisons with pure OMA}
Since the proposed hybrid NOMA scheme ensures that $U_m$ has the same performance as in OMA, the rest of the paper will focus on the performance of $U_n$ to unveil the advantages of hybrid NOMA. To this end, the following data rate maximization problem is first formulated:
\SubEquL{Primal_Prob}{
  \max_{\beta_1,\beta_2} & \quad\quad R_n\\
   s.t.   & \quad\quad \beta_1+\beta_2 \leq \eta,
  \\
  &\quad\quad \beta_1\rho_n |h_n|^2\leq \tau_m,
}
where (\ref{Primal_Prob}b) constrains the energy consumption of hybrid NOMA. Note that
$\eta$ is a constant and satisfies $\eta\leq 1$, which guarantees that
the energy consumption of hybrid NOMA cannot exceed that of OMA. (\ref{Primal_Prob}c)
is considered to ensure the success of decoding $U_m$'s signal as stated in Section II.

It can be easily concluded from problem (\ref{Primal_Prob}) that the considered hybrid NOMA
is a general form, since both pure NOMA ($\beta_2=0$) and OMA ($\beta_1=0$) can be treated as
special cases of hybrid NOMA.

The optimal solution of problem (\ref{Primal_Prob}) can be obtained in a closed-form, as highlighted in the following
lemma.
\begin{Lemma}
The optimal solution of problem (\ref{Primal_Prob}) can be expressed as follows:
\Equ{\beta_1^*=\left\{
               \begin{array}{ll}
                 0, & {\tau_m=0}, \\
                 \frac{\eta}{2}, & {\tau_m>0\text{ and }\frac{\tau_m}{\rho_n|h_n^2|}\geq \frac{\eta}{2}},\\
                 \frac{\tau_m}{\rho_n|h_n^2|}, & {\tau_m>0\text{ and }\frac{\tau_m}{\rho_n|h_n^2|}< \frac{\eta}{2}},\\
               \end{array}
             \right.
}
and $\beta_2^*=\eta-\beta_1^*$.
\end{Lemma}
\begin{IEEEproof}
Please refer to Appendix A.
\end{IEEEproof}

\begin{Remark}
From the results in Lemma $1$, the achievable data rate of the considered hybrid NOMA, denoted by $R_n^*$,  can be straightforwardly obtained as follows:
\begin{itemize}
\item When $\tau_m=0$,
 \Equ{R_n^*=\log(1+\eta\rho_n|h_n|^2)}
 For this case, $\tau_m=0$ means that $U_m$ does not have the ability to accommodate inter-user interferences, and hence $U_n$ is prohibited from transmitting any signal in $U_m$'s time slot. Thus, $U_n$ can only transmit by using its own time slot by using pure OMA. In this case, the data rate achieved by hybrid NOMA equals to that achieved by pure OMA only if $\eta=1$.

\item When $\tau_m>0$,
\Equ{R_n^*=2\log(1+\frac{\eta}{2}\rho_n|h_n|^2),}
for $\frac{\tau_m}{\rho_n|h_n^2|}\geq \frac{\eta}{2}$, and
\Equ{R_n^*=\log(1+\tau_m)+\log(1+\eta\rho_n|h_n|^2-\tau_m),}
for $\frac{\tau_m}{\rho_n|h_n^2|}< \frac{\eta}{2}$. For this case, $\tau_m>0$ means that
$U_m$ has the ability to accommodate some interferences from other users. An interesting and important
observation is:  when $\tau_m>0$, $U_n$ always prefer to occupying $U_m$'s slot to transmit partially, i.e., $\beta_1^*>0$.
In other words, hybrid NOMA achieves a higher data rate than pure NOMA and pure OMA when $\tau_m>0$.
\end{itemize}
\end{Remark}

Next, we will show the superior performance of hybrid NOMA
in terms of energy efficiency compared to OMA.  From Lemma $1$, it can be found that the energy consumption of the considered hybrid NOMA scheme is $\eta T \rho_n$ ($0<\eta \leq 1$), which can be lower than
$T \rho_n$ of the pure OMA scheme for $\eta<1$. Thus, an interesting question is whether
hybrid NOMA can achieve a higher data rate with $\eta<1$ compared to a pure OMA scheme which uses full transmit power and has an achievable data rate as shown in (1).  By comparing $R_n^*$ and $R_n^{\text{OMA}}$ with some algebraic manipulations, the following corollary can be obtained.

\begin{corollary}
For a given $\eta$,  the maximum achievable data rate of hybrid NOMA
is strictly larger than that of OMA which utilizes full transmit power $\rho_n$, i.e., $R_n^*>R_n^{\text{OMA}}$, if and only if:
\Equ{\frac{\tau_m}{\rho_n|h_n^2|}\geq \frac{\eta}{2}, \text{ and } |h_n|^2>\frac{4(1-\eta)}{\eta^2\rho_n},}
or
\Equ{
0<\frac{\tau_m}{\rho_n|h_n^2|}< \frac{\eta}{2}, \tau_m\geq \frac{1}{\eta}-1,
\text{and } |h_n|^2>\frac{\tau_m^2}{\rho_n\left(\eta+\tau_m\eta-1\right)}.
}

\end{corollary}

\begin{Remark}
When $\eta=1$, it can be straightforwardly  observed that $R_n^*>R_n^{\text{OMA}}$ always holds, as long as $\tau_m>0$.
\end{Remark}

Corollary $1$ gives the conditions under which  hybrid NOMA can achieve a higher instantaneous data rate
than pure NOMA by consuming less energy. Furthermore, it is interesting to investigate how likely these conditions
occur  by considering the randomness of the channel gains. Particularly, define the following probability:
\Equ{
 P_n^w=\text{Pr}\left(R_n^*\leq R_n^{\text{OMA}}\right).
}
The expression for $P_n^w$ can be obtained as highlighted in the following lemma.

\begin{Lemma}
For a given $\eta$, $0<\eta\leq1$, the probability that $R_n^*\leq R_n^{\text{OMA}}$ can be expressed
as:
\Equ{P_n^w=1-e^{-\frac{(2-\eta)\epsilon_0}{\eta\rho_m}-\frac{4(1-\eta)}{\eta^2\rho_n}}
-\int_{\frac{\epsilon_0}{\eta\rho_m}}^{\frac{(2-\eta)\epsilon_0}{\eta\rho_m}}
  e^{-y}e^{-\frac{(\frac{\rho_m}{\epsilon_0}y-1)^2}{\rho_n\left(\frac{\eta\rho_m}{\epsilon_0}y-1\right)}}\,dy
}
\end{Lemma}
\begin{IEEEproof}
Please refer to Appendix B.
\end{IEEEproof}

\begin{Remark}
By taking derivatives of $P_n^w$ with respect to $\rho_n$ and $\rho_m$, respectively, it can be found that
$P_n^w$ decreases with $\rho_n$ and $\rho_m$, respectively.
\end{Remark}

Based on Lemma $2$, the following Lemma can be obtained, which characterizes the asymptotic performance of $P_n^w$
in the high SNR regime.

\begin{Lemma}
For a given $\eta$, $0<\eta\leq1$, when $\rho_n \rightarrow \infty $ and $\rho_m \rightarrow \infty$,  the probability that $R_n^*\leq R_n^{\text{OMA}}$ can be approximated as follows:
\Equ{\label{App_P_n^w}
P_n^w\approx \frac{\epsilon_0}{\eta\rho_m}+\frac{4(1-\eta)}{\eta^2\rho_n}
}
\end{Lemma}
\begin{IEEEproof}
Please refer to Appendix C.
\end{IEEEproof}

\begin{Remark}\label{P_zero}
From Lemma $3$, it can be straightforwardly observed that $P_n^w$ can approach zero when $\rho_n \rightarrow \infty $ and $\rho_m \rightarrow \infty$, for any given $\eta$, $0<\eta\leq1$. The importance of this observation is: even with less energy consumption (when $\eta<1$), hybrid NOMA can almost surely achieve a higher instantaneous
data rate than pure OMA in the high SNR regime, which indicates the superior energy efficiency of hybrid NOMA.
\end{Remark}

\begin{Remark}\label{P_constant}
Note that $P_n^w$ can approach zero only when both $\rho_n$ and $\rho_m$ becomes sufficiently large. Specifically, by following the similar methods in Appendix C, it can be proved that:
when $\rho_n$ or $\rho_m$ is a constant, while the other goes infinity, $P_n^w$ approaches a constant, as listed in the following:
\begin{itemize}
\item when $\rho_n$ is a constant and  $\rho_m \rightarrow \infty $, $P_n^w$ can be approximated as:
\Equ{
P_n^w\approx 1-e^{-\frac{4(1-\eta)}{\eta^2\rho_n}}.
}
  \item when $\rho_m$ is a constant and  $\rho_n \rightarrow \infty $, $P_n^w$ can be approximated as:
  \Equ{\label{rho_m_infty}
P_n^w\approx 1-e^{-\frac{\epsilon_0}{\eta\rho_m}}.
}
It is interesting in (\ref{rho_m_infty}) that, when $\rho_m$ is a constant $P_n^w$ is limited not only  by $\rho_m$, but also by $\eta$.
\end{itemize}

\end{Remark}

\section{Numerical Results}
In this section, numerical results are presented to verify the developed analytical results and also demonstrate the superior performance of the proposed hybrid NOMA scheme.

\begin{figure}[!t]
  \centering
      \setlength{\abovecaptionskip}{0em}   % ����ͼƬ������ͼƬ����
      \setlength{\belowcaptionskip}{-1em}   % ����ͼƬ���������ľ���
  \includegraphics[width=2.8in]{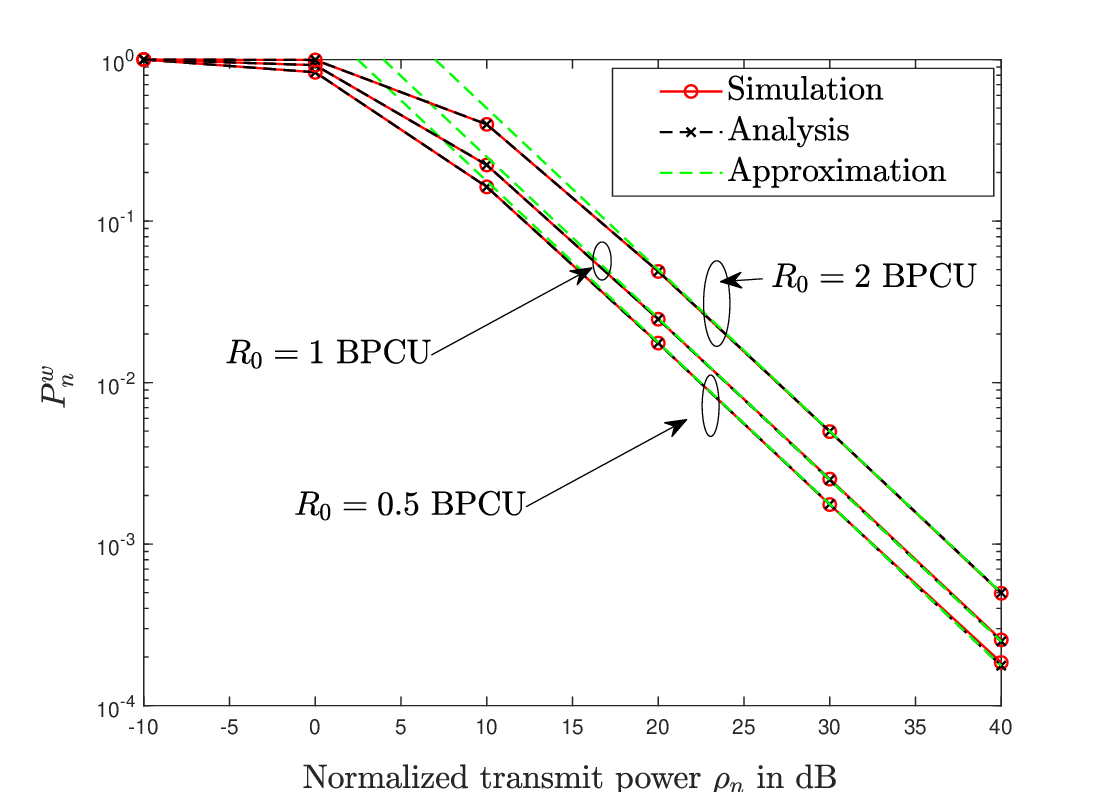}\\
  \caption{The probability of the event that $R_n^*\leq R_n^{\text{OMA}}$. $\eta=0.8$, $\rho_n=\rho_m$.}\label{accuracy}
\end{figure}

Fig. \ref{accuracy} shows the probability of the event that $R_n^*\leq R_n^{\text{OMA}}$, i.e., $P_n^w$.
Note that ``BPCU'' stands for ``bits per channel use''.
It is worth pointing out that the energy consumption of the hybrid NOMA is $T\eta\rho_n$, while the energy consumption of the pure OMA is $T\rho_n$.
The analytical results are  based on Lemma $2$, and the approximations are based on Lemma $3$.
From the figure, it can be seen that the analytical results are in perfect agreement with the simulations, which verifies
the accuracy of the analysis. It can also be seen that the curves of the approximated results
overlap with those of the simulation results. Therefore, the approximations are accurate in the high SNR regime.

\begin{figure}[!t]
  \centering
      \setlength{\abovecaptionskip}{0em}   % ����ͼƬ������ͼƬ����
      \setlength{\belowcaptionskip}{-2em}   % ����ͼƬ���������ľ���
  \includegraphics[width=2.8in]{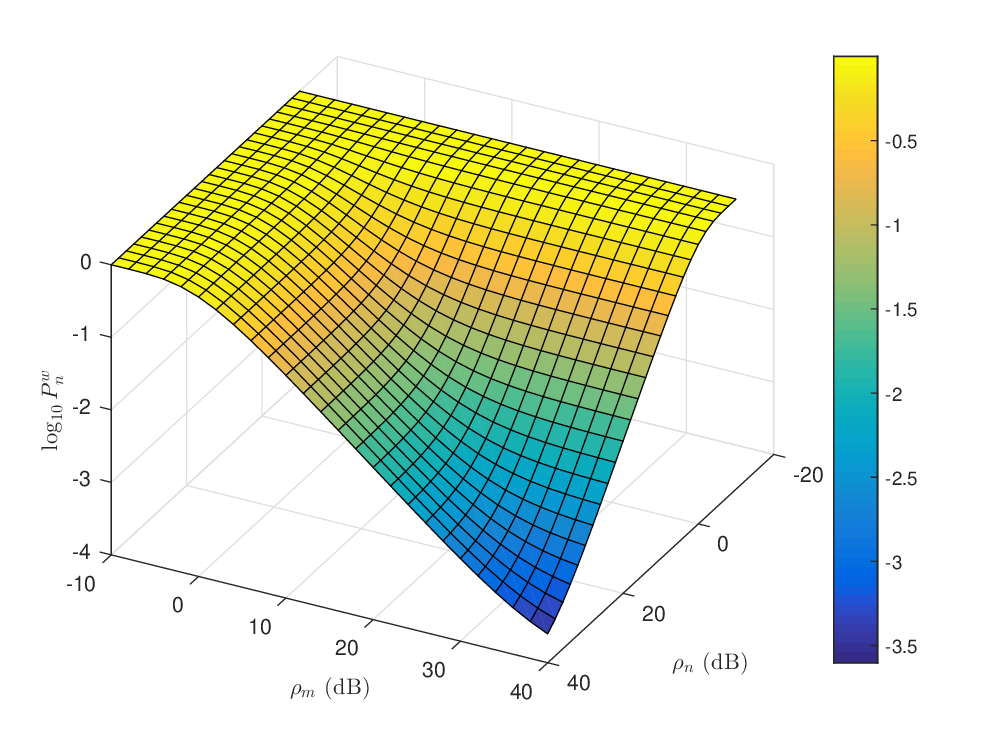}\\
  \caption{The probability of the event that $R_n^*\leq R_n^{\text{OMA}}$. $\eta=0.8$, $R_0=1$ BPCU.}\label{surf}
\end{figure}

Fig. \ref{surf} shows how $P_n^w$ varies with respect to $\rho_n$ and $\rho_m$, which is based on Lemma $2$. From the figure,
it can be observed that  when both $\rho_n$ and $\rho_m$ are sufficiently large, $P_n^w$ approaches zero, which is consistent with the conclusions of Remark \ref{P_zero}.
From Fig. \ref{surf}, it can also be seen that for fixed $\rho_n$, $P_n^w$ decreases as $\rho_m$ increases.
However, there's a saturation  for  $P_n^w$, indicating that if $\rho_n$ is bounded, further
increasing $\rho_m$ won't bring $P_n^w$ to zero. Similar conclusions can be observed when $\rho_n$ is fixed.
These observations are consistent with the statement discussed in Remark \ref{P_constant}.

\begin{figure}[!t]
  \centering
      \setlength{\abovecaptionskip}{0em}   % ����ͼƬ������ͼƬ����
      \setlength{\belowcaptionskip}{-2em}   % ����ͼƬ���������ľ���
  \includegraphics[width=2.8in]{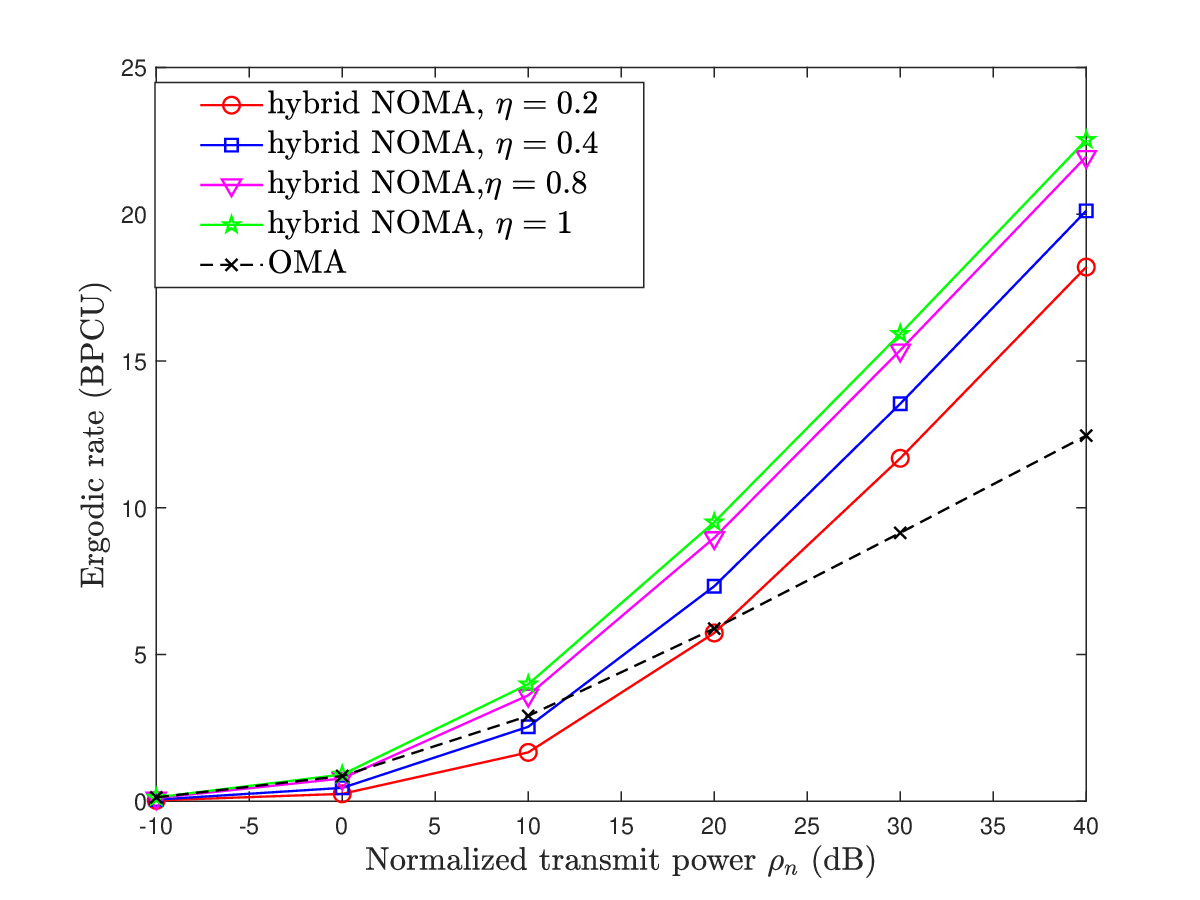}\\
  \caption{Ergodic data rate achieved by hybrid NOMA and pure OMA. $\rho_n=\rho_m$, $R_0=1$ BPCU.}\label{ergodic_rate}
\end{figure}

Fig. \ref{ergodic_rate} compares the ergodic rates achieved by hybrid NOMA and pure OMA.
Note that the ergodic data rate is obtained by averaging over $10^7$ random channel realizations.
From the figure, it can be observed that when $\eta<1$, the ergodic data rate achieved by
hybrid NOMA is lower than that of pure OMA at low SNRs. However, as $\rho_n$ increases,
the ergodic data rate achieved by hybrid NOMA can outperform that of pure OMA, and the gap increases
with $\rho_n$. Thus, it can be concluded that hybrid NOMA can achieve a higher ergodic data rate while consuming less power than pure OMA in the high SNR regime.

\section{Conclusions}
In this paper, the achievable data rate of hybrid NOMA has been obtained by optimizing the power allocation. The conditions under which hybrid NOMA can achieve a higher instantaneous data rate with less power consumption compared to pure OMA have been established. Furthermore, it has been proved that the probability that hybrid NOMA can achieve a higher instantaneous data rate with less power consumption (for any given $\eta<1$) compared to OMA approaches one in the high SNR regime. Numerical results are also provided to show that the ergodic data rate achieved by hybrid NOMA with lower energy consumption is greater than that of pure OMA.

\appendices
\section{Proof for Lemma 1}
It is straightforward  to prove that the equality in constraint (\ref{Primal_Prob}b)  must hold for the optimal solution of
problem (\ref{Primal_Prob}), i.e., $\beta_2=\eta-\beta_1$. Thus, the primal problem is equivalent to the following:
\SubEquL{equi_Prob_1}{
  \max_{\beta_1} &\log\left(1+\beta_1\rho_n|h_n|^2\right)+ \log\left(1+(\eta-\beta_1)\rho_n|h_n|^2\right)\\
   s.t.  & \quad (\ref{Primal_Prob}c). \notag
}

Then, it can be observed that finding the optimal $\beta_1$ is equivalent to solving the following optimization problem:
\SubEquL{equi_Prob_2}{
  \max_{\beta_1} &-\rho_n^2|h_n|^4\beta_1^2+\eta\rho_n^2|h_n|^4\beta_1+1+\eta\rho_n|h_n|^2\\
   s.t.  & \quad (\ref{Primal_Prob}c). \notag
}
Note that the objective function of problem (\ref{equi_Prob_2}) is a quadratic function of $\beta_1$, and
the optimal solution for  (\ref{equi_Prob_2}a) without any constraint is $\frac{\eta}{2}$. Therefore, the optimal solution of the primal problem can be determined by finding the feasible solution in (\ref{Primal_Prob}c) which is nearest to $\frac{\eta}{2}$, and the proof for Lemma $1$ can be complete.

\section{Proof for Lemma 2}
By considering the conditions shown in Corollary 1, $P_n^w$ can be written as follows:
\Equ{P_n^w=&\underset{P_1}{\underbrace{\text{Pr}\left(\tau_m=0\right)}}+
\underset{P_2}{\underbrace{\text{Pr}\left(\tau_m>0, \frac{\tau_m}{\rho_n|h_n|^2}\geq \frac{\eta}{2}\right)}} \\\notag
&+\underset{P_{3,1}}{\underbrace{\text{Pr}\left(\tau_m>0, \frac{\tau_m}{\rho_n|h_n|^2}<\frac{\eta}{2}, \tau_m\leq \frac{1}{\eta}-1\right)}} \\\notag
&+\text{Pr}\left(\tau_m>0, \frac{\tau_m}{\rho_n|h_n|^2}<\frac{\eta}{2}, \tau_m > \frac{1}{\eta}-1,\right. \\\notag&\quad\quad\quad  \underset{P_{3,2}}{\underbrace{\left.|h_n|^2\leq \frac{\tau_m^2}{\rho_n(\eta+\tau_m\eta-1)}\right)}}.
}
Thus,  the remaining task is to calculate the expressions for $P_1$, $P_2$, $P_{3,1}$ and $P_{3,2}$.

The expression for $P_1$ can be obtained as follows:
\Equ{
P_1&=\text{Pr}\left(\!|h_m|^2\leq \frac{\epsilon_0}{\rho_m}\!\right)
=\int_{0}^{\frac{\epsilon_0}{\rho_m}}e^{-x}\,dx=1-e^{-\frac{\epsilon_0}{\rho_m}}.
}

$P_2$ can be rewritten as
\Equ{P_2=&\text{Pr}\left(\!|h_m|^2\!>\!\frac{\epsilon_0}{\rho_m}, |h_n|^2\leq\min\left\{\frac{2\tau_m}{\eta\rho_n},\frac{4(1\!-\!\eta)}{\eta^2\rho_n}\right\}\right)\\\notag
=&\underset{P_{2,1}}{\underbrace{\text{Pr}\left(|h_m|^2>\frac{\epsilon_0(2-\eta)}{\rho_m\eta}, |h_n|^2\leq\frac{4(1-\eta)}{\eta^2\rho_n}\right)}}+\\\notag
&\underset{P_{2,2}}{\underbrace{\text{Pr}\left(\frac{\epsilon_0}{\rho_m}<|h_m|^2\leq\frac{\epsilon_0(2-\eta)}{\rho_m\eta}, |h_n|^2\leq\frac{2\tau_m}{\eta\rho_n}\right)}}
}
By noting that the probability density functions (PDF) for $|h_n|^2$ and $|h_m|^2$ are
\Equ{f_{|h_n|^2}(x)= e^{-x}, f_{|h_m|^2}(y)= e^{-y},}
$P_{2,1}$ and $P_{2,2}$  can be obtained as follows:
\begin{align}
 P_{2,1}&=\int_{\frac{\epsilon_0(2-\eta)}{\rho_m\eta}}^{+\infty}\int_{0}^{\frac{4(1-\eta)}{\eta^2\rho_n}}e^{-x}e^{-y}\,dxdy\\\notag
         &=e^{-\frac{\epsilon_0(2-\eta)}{\rho_m\eta}}\left(1-e^{-\frac{4(1-\eta)}{\eta^2\rho_n}}\right)
\end{align}
\Equ{
 P_{2,2}&=\int_{\frac{\epsilon_0}{\rho_m}}^{\frac{\epsilon_0(2-\eta)}{\rho_m\eta}}\int_{0}^{\frac{2\rho_my-2\epsilon_0}{\eta\rho_n\epsilon_0}}
 e^{-x}e^{-y}\,dxdy\\\notag
 &=e^{-\frac{\epsilon_0}{\rho_m}}-e^{-\frac{\epsilon_0(2-\eta)}{\rho_m\eta}}\\\notag
 &-\frac{\eta\rho_n\epsilon_0e^{\frac{2}{\eta\rho_n}}}{2\rho_m+\eta\rho_n\epsilon_0}
 \left(e^{-\frac{2}{\eta\rho_n}-\frac{\epsilon_0}{\rho_m}}-e^{-\frac{(2-\eta)\epsilon_0}{\eta\rho_m}-\frac{4-2\eta}{\eta^2\rho_n}}\right)
}
Thus, the expression for $P_2$ can be obtained by summing $P_{2,1}$ and $P_{2,2}$.

$P_{3,1}$ can calculated as follows:
\Equ{
 P_{3,1}&=\text{Pr}\left(\frac{\epsilon_0}{\rho_m}<|h_m|^2\leq \frac{\epsilon_0}{\eta\rho_m}, |h_n|^2>\frac{2\tau_m}{\eta\rho_n}\right)\\\notag
  &=\int_{\frac{\epsilon_0}{\rho_m}}^{\frac{\epsilon_0}{\eta\rho_m}}\int_{\frac{2\rho_my-2\epsilon_0}{\eta\rho_n\epsilon_0}}^{+\infty}e^{-x}e^{-y}\,dxdy\\\notag
  &=\frac{\eta\rho_n\epsilon_0e^{\frac{2}{\eta\rho_n}}}{2\rho_m+\eta\rho_n\epsilon_0}
  \left(e^{-\frac{2}{\eta\rho_n}-\frac{\epsilon_0}{\rho_m}}-e^{-\frac{2}{\eta^2\rho_n}-\frac{\epsilon_0}{\eta\rho_m}}\right)
}

By noting that $\eta\leq 1$, $P_{3,2}$ can rewritten as follows:
\Equ{\label{P_{3,2}}
P_{3,2}=&\text{Pr}\left(|h_m|^2>\frac{\epsilon_0}{\eta\rho_m}, |h_n|^2>\frac{2\tau_m}{\eta\rho_n},\right.\\\notag
&\quad\quad\left. |h_n|^2\leq \frac{\tau_m^2}{\rho_n(\eta+\tau_m\eta-1)}\right),
}
It is worth pointing out that there is an implicit condition in (\ref{P_{3,2}}) that
\Equ{
 \frac{2\tau_m}{\eta\rho_n}<\frac{\tau_m^2}{\rho_n(\eta+\tau_m\eta-1)}.
}
which can be simplified to
$
|h_m|^2<\frac{(2-\eta)\epsilon_0}{\eta\rho_m}.
$
Thus, $P_{3,2}$ should be expressed as:
\Equ{
P_{3,2}=&\text{Pr}\left(\frac{\epsilon_0}{\eta\rho_m}<|h_m|^2<\frac{(2-\eta)\epsilon_0}{\eta\rho_m}, \right.\\\notag
&\quad\quad\left.\frac{2\tau_m}{\eta\rho_n}<|h_n|^2\leq \frac{\tau_m^2}{\rho_n(\eta+\tau_m\eta-1)}\right)
}
By taking the PDFs for $|h_n|^2$ and $|h_m|^2$ into the above equation, $P_{3,2}$ can be expressed as:
\Equ{
P_{3,2}&=\int_{\frac{\epsilon_0}{\eta\rho_m}}^{\frac{(2-\eta)\epsilon_0}{\eta\rho_m}}\int_{\frac{2\tau_m}{\eta\rho_n}}^{\frac{\tau_m^2}{\rho_n(\eta+\tau_m\eta-1)}}e^{-x}e^{-y}\,dxdy\\\notag
       &=\int_{\frac{\epsilon_0}{\eta\rho_m}}^{\frac{(2-\eta)\epsilon_0}{\eta\rho_m}}e^{-y}e^{-\frac{2\rho_my-2\epsilon_0}{\eta\rho_n\epsilon_0}}\,dy\\\notag
       &\quad -\int_{\frac{\epsilon_0}{\eta\rho_m}}^{\frac{(2-\eta)\epsilon_0}{\eta\rho_m}}e^{-y}e^{-\frac{(\frac{\rho_m}{\epsilon_0}y-1)^2}{\rho_n\left(\frac{\eta\rho_m}{\epsilon_0}y-1\right)}}\,dy\\\notag
      &= \frac{\eta\rho_n\epsilon_0e^{\frac{2}{\eta\rho_n}}}{2\rho_m+\eta\rho_n\epsilon_0}
 \left(e^{-\frac{2}{\eta^2\rho_n}-\frac{\epsilon_0}{\eta\rho_m}}-e^{-\frac{(2-\eta)\epsilon_0}{\eta\rho_m}-\frac{4-2\eta}{\eta^2\rho_n}}\right)\\\notag
 &\quad -\int_{\frac{\epsilon_0}{\eta\rho_m}}^{\frac{(2-\eta)\epsilon_0}{\eta\rho_m}}e^{-y}e^{-\frac{(\frac{\rho_m}{\epsilon_0}y-1)^2}{\rho_n\left(\frac{\eta\rho_m}{\epsilon_0}y-1\right)}}\,dy
}
By summing up $P_1$, $P_2$, $P_{3,1}$ and $P_{3,2}$ and making some algebraic manipulations, the expression for
$P_n^w$ can be obtained and the proof is complete.

\section{Proof for Lemma 3}
Denote $P_n^w$ by $P_n^w=G+Q$,  where
\Equ{G=1-e^{-\frac{(2-\eta)\epsilon_0}{\eta\rho_m}-\frac{4(1-\eta)}{\eta^2\rho_n}},
}
and
\Equ{Q=-\int_{\frac{\epsilon_0}{\eta\rho_m}}^{\frac{(2-\eta)\epsilon_0}{\eta\rho_m}}
  e^{-y}e^{-\frac{(\frac{\rho_m}{\epsilon_0}y-1)^2}{\rho_n\left(\frac{\eta\rho_m}{\epsilon_0}y-1\right)}}\,dy.
}
By applying the Taylor series $e^{-x}\approx 1-x$ ($x\rightarrow 0$), $G$ can be approximated as follows:
\Equ{
G&\approx1-\left(1-\frac{(2-\eta)\epsilon_0}{\eta\rho_m}\right)\left(1-\frac{4(1-\eta)}{\eta^2\rho_n}\right)\\\notag
&=\frac{(2-\eta)\epsilon_0}{\eta\rho_m}+\frac{4(1-\eta)}{\eta^2\rho_n}-\frac{4(1-\eta)(2-\eta)\epsilon_0}{\eta^3\rho_m\rho_n}\\\notag
&\overset{(a)}{\approx} \frac{(2-\eta)\epsilon_0}{\eta\rho_m}+\frac{4(1-\eta)}{\eta^2\rho_n},
}
where the step (a)  follows from the fact that the last term is a higher order infinitesimal quantity compared to the other two terms and hence can be neglected.

Let $t=\frac{\rho_m}{\epsilon_0}y$, $Q$ can be written as
\Equ{
Q=-\int_{\frac{1}{\eta}}^{\frac{2-\eta}{\eta}}\frac{\epsilon_0}{\rho_m}e^{-\frac{\epsilon_0}{\rho_m}t}
   e^{-\frac{(t-1)^2}{\rho_n(\eta t-1)}}\,dt.
}
By taking  Taylor series, $Q$ can be approximated as follows:
\Equ{
Q&\approx-\int_{\frac{1}{\eta}}^{\frac{2-\eta}{\eta}}\frac{\epsilon_0}{\rho_m}\left(1-\frac{\epsilon_0}{\rho_m}t\right)
  \left(1-\frac{(t-1)^2}{\rho_n(\eta t-1)}\right)\,dt\\\notag
  &\approx -\int_{\frac{1}{\eta}}^{\frac{2-\eta}{\eta}}\frac{\epsilon_0}{\rho_m}\,dt=-\frac{(1-\eta)\epsilon_0}{\eta\rho_m}
}
Thus, the approximate for $P_n^w$ in (\ref{App_P_n^w}) can be obtained
and the proof is complete.
\bibliographystyle{IEEEtran}
\bibliography{IEEEabrv,ref}
\end{document}